\documentclass[twocolumn]{aa}
\usepackage{graphicx}
\usepackage{txfonts}
\usepackage{subcaption}

\usepackage[dvipsnames]{xcolor}
\usepackage{hyperref}
\colorlet{mylinkcolor}{BlueViolet}
\colorlet{mycitecolor}{BlueViolet}
\colorlet{myurlcolor}{BlueViolet}
\newcommand\myshade{85}

\hypersetup{
  linkcolor  = mylinkcolor!\myshade!black,
  citecolor  = mycitecolor!\myshade!black,
  urlcolor   = myurlcolor!\myshade!black,
  colorlinks = true,
}

\begin{document} 
    
\title{Spectroscopic Transit Search: a self-calibrating method for detecting planets around bright stars}

\author{Lennart van Sluijs
          \inst{1}
          \and
          Ernst de Mooij
          \inst{2}
          \and
          Matthew Kenworthy
          \inst{1}
          \and
          Maggie Celeste
          \inst{2}
          \and
          Matthew J. Hooton
          \inst{3}
          \and
          Eric E. Mamajek
          \inst{4, 5}
          \and
          Brigitta Sip\H{o}cz
          \inst{6}
          \and
          Ignas. A. G. Snellen
          \inst{1}
          \and
          Andrew R. Ridden-Harper
          \inst{7}
          \and
          Paul A. Wilson
          \inst{8, 9}
      }

\institute{Leiden Observatory, P.O. Box 9513, NL-2300 RA Leiden, The Netherlands\\
              \email{vansluijs@strw.leidenuniv.nl}
         \and
School of Physical Sciences, and Centre for Astrophysics and Relativity, Dublin City University, Glasnevin, Dublin 9, Ireland
        \and
        Astrophysics Research Centre, School of Mathematics and Physics, Queen’s University Belfast, Belfast BT7 1NN, UK
\and
Jet Propulsion Laboratory,
California Institute of Technology
4800 Oak Grove Dr., Pasadena, CA 91109, USA
\and
Department of Physics and Astronomy, University of Rochester, 500 Wilson Blvd., Rochester, NY 14627, USA
        \and
Institute of Astronomy, University of Cambridge, Madingley Road, Cambridge, CB3 0HA, UK
        \and
        Department of Astronomy, Cornell University, Ithaca, New York 14853, USA
        \and
Department of Physics, University of Warwick, Coventry CV4 7AL, UK
        \and
Centre for Exoplanets and Habitability, University of Warwick, Coventry CV4 7AL, UK
            }

   \date{Received XX XX, XXXX accepted YY YY, YYYY}

 
  \abstract
   {}
   {We search for transiting exoplanets around the star $\beta$ Pictoris using high resolution spectroscopy and Doppler imaging that removes the need for standard star observations. These data were obtained on the VLT with UVES during the course of an observing campaign throughout 2017 that monitored the Hill sphere transit of the exoplanet $\beta$ Pictoris b.}
   {We utilize line profile tomography as a method for the discovery of transiting exoplanets. By measuring the exoplanet distortion of the stellar line profile, we remove the need for reference star measurements. We demonstrate the method with white noise simulations, and then look at the case of $\beta$ Pictoris, which is a $\delta$ Scuti pulsator. We describe a method to remove the stellar pulsations and perform a search for any transiting exoplanets in the resultant data set. We inject fake planet transits with varying orbital periods and planet radii into the spectra and determine the recovery fraction.}
   {In the photon noise limited case we can recover planets down to a Neptune radius with an $\sim$80\% success rate, using an 8\,m telescope with a $R\sim 100,000$ spectrograph and 20 minutes of observations per night. The pulsations of $\beta$ Pictoris limit our sensitivity to Jupiter-sized planets, but a pulsation removal algorithm improves this limit to Saturn-sized planets. We present two planet candidates, but argue that their signals are most likely caused by other phenomena.}
   {We have demonstrated a method for searching for transiting exoplanets that (i) does not require ancillary calibration observations, (ii) can work on any star whose rotational broadening can be resolved with a high spectral dispersion spectrograph and (iii) provides the lowest limits so far on the radii of transiting Jupiter-sized exoplanets around $\beta$ Pictoris with orbital periods from 15 days to 200 days with >50$\%$ coverage.}


\maketitle

\section{Introduction}
\label{s:intro}
A majority of the exoplanets discovered to date has been through the simultaneous photometric monitoring of several thousands of stars and looking for the decrement in stellar flux as a companion transits the stellar disk.
Several ground based photometric surveys, such as TrES \citep{Alonso04}, XO \citep{McCullough05}, SuperWASP \citep{Pollacco06,Butters10}, HATNet \citep{Bakos07}, and NGTS \citep{Wheatley18} and several space missions, such as Kepler \citep{Borucki10} and CoRoT have been successful in detecting new exoplanets.
Over two thousand transiting exoplanets have now been detected, and follow up missions now include the TESS \citep{Ricker14} and PLATO \citep{Rauer14} space missions.
Transmission spectroscopy on these transiting exoplanets enables the characterization and detection of their atmospheres.
The brighter the star, the higher the signal to noise of the resultant exoplanet atmospheric spectrum \citep{Seager10}.
Finding the brightest star with a transiting exoplanet, therefore, is an important science goal that is being led from the ground by the WASP \citep{Anderson18}, KELT \citep{Lund17} and MASCARA \citep{Talens17a, Talens18} surveys, with MASCARA monitoring the brightest stars up to $V = 4$ \citep{Talens17}.
From space it is being led by TESS \citep{Ricker14} which also goes as bright as approximately $V = 4$.
Ironically, despite the abundance of photons the brightest $V < 4$ stars in the sky are not monitored by current transit surveys.
This is largely because of the significant challenges in calibrating photometry of bright stars in wide field surveys, detailed in \citet{Talens17}. This is mostly due to the significantly different light paths from equally bright stars through the optics of a telescope and additionally, for ground based telescopes, through the Earth's atmosphere. 
The limited field of view of larger telescopes means that it is very difficult to find a bright photometric standard with which to calibrate bright $V < 4$ star transit observations.
Therefore, in this paper we present an alternative method for the detection of a transiting exoplanet that does not require a calibration star and thus can be utilized to survey the brightest $V < 4$ stars in the sky, only requiring they have sufficiently fast rotations.
We look for the distortion of the rotationally broadened chromospheric stellar lines as a planet transits the stellar disk, also known as the Rossiter-McLaughlin (RM) effect.
This technique is commonly used to determine the spin-orbit alignment of exoplanet and host stars during known transits \citep[for an extensive overview, see][]{Triaud17}.
Here we carry out a blind search for a transiting exoplanet using multi-epoch high spectral resolution observations of a bright star, calibrated using only the target star spectra.
Many bright stars in the night sky are intermediate-mass main-sequence stars.
Their fast rotations \citep[e.g.][and references therein]{Gray05} broaden the chromospheric lines which limit their radial velocity sensitivity, and early type stars have typically far fewer absorption lines to provide a precise determination of their radial velocity.
In this paper we use $\beta$ Pictoris, a typical fast-rotating bright ($V < 4$) star, to investigate the feasibility of this method.
The nearby \citep{vanLeeuwen07} bright young \citep[$\sim$23 Myr;][]{Mamajek14} A6V star $\beta$ Pictoris has both a debris disk and at least one giant planet, $\beta$ Pictoris b \citep{Lagrange09,Lagrange18a} in orbit around it.
Both the disk and the planet are seen nearly edge-on with a very high inclination of $> 89\degr$ \citep{Wang16}.
In 2017 the Hill sphere of the planet moved in front of the star, taking approximately 200 days to move across in a chord that brought the line of sight of the star to within 20\% of the Hill sphere radius.
A comprehensive campaign of photometric and spectroscopic observations were taken over this period searching for circumplanetary material \citep{2017NatAs...1E..99K}.
Due to the abundance of high-resolution spectra available as part of this campaign (PI: E. de Mooij), $\beta$ Pictoris is an excellent target to study the feasibility of our method.
Firstly, in Section~\ref{s:principle} we outline the method by studying its potential by simulating observations of a transiting companion around a fast rotating $V\sim4$ bright star.
Secondly, in Section~\ref{s:datasensitivity}, we apply our method to real data of $\beta$ Pictoris as a case study.
The results are discussed in Section~\ref{s:discussion} followed by our conclusions and future prospects in Section~\ref{s:conclusions}.

\section{Principle of the method}\label{s:principle}

In this section we demonstrate the application of the RM effect to the discovery of new planets.
For this we first explain how the RM effect is modeled in the next subsection and then highlight its ability to recover planetary signals using a set of white noise simulations of a bright (V$\sim$4) star with a high-resolution spectrograph on an 8 m telescope.

\subsection{RM model}
\label{ss:rmmodel}
Our RM model is based on the model used by~\citet{DeMooij17}, and uses a grid-based method to calculate the line-profile of a rotating star.
For the model we assume solid body rotation, quadratic limb darkening and no gravity darkening.
The intrinsic line-profile, $F_{\rm{ij}}(v)$, at a pixel location $(i, j)$, is modeled as a Gaussian with a line-depth $A$, a width given by the Full Width at Half Maximum, FWHM. For each pixel, the line-profile is centred on a radial velocity $v_{\rm{rot, ij}}$, due to the stellar rotation at that position. The planet is modeled as a black disk at position $(i_{\rm{p}}, 0)$, assuming an orbit parallel to the x-axis and a projected spin-orbit misalignment $\lambda = 0\degr$ with impact parameter $b = 0$.
For all our simulations, we use a grid of 1025 by 1025 spatial pixels for the calculations of the spectrum with a stellar radius of 510 pixels.
To reduce the impact of aliasing effects, especially for smaller planets and at ingress and egress, both the stellar intensity map and the planet map are initially calculated on a grid that is over-sampled by a factor of 10 in both directions, and rebinned to 1025 by 1025 pixels before calculating the final spectrum. 
This spectrum is calculated on a velocity grid of $3 \ \rm{km/s}$ steps.

\subsection{White noise simulation}
\label{ss:wnsim}

\begin{figure}
    \centering
    \includegraphics[width= \columnwidth]{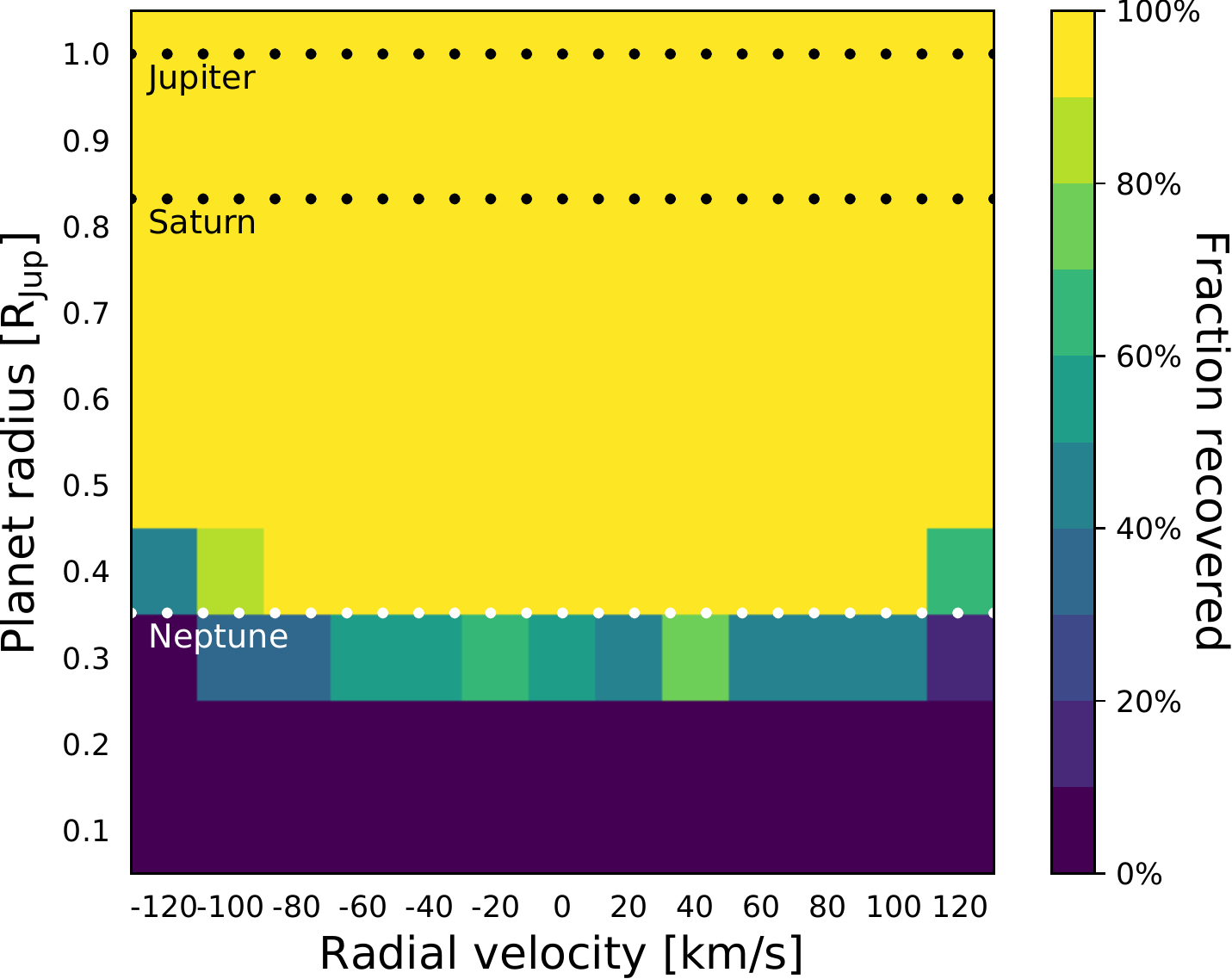}
    \caption{Recovery fraction of transit injections in 152 nights for our white noise simulation. Exoplanet signals with a SNR $> 3.0$ above other nights are considered recovered. We assume an impact parameter $b = 0$ and a star observed at SNR = 1200/pix.
    The radii of Jupiter, Saturn and Neptune are indicated by the black and white dashed lines.
    As $v_{\rm{eq}} = 130 \ \rm{km/s}$, radial velocities outside of our range are (close-to) zero.
    The effects of limb darkening are seen in the bottom rows as lower recovery fractions for higher radial speeds.}
    \label{fig:wnsen}
\end{figure}

We demonstrate the method by considering photon shot noise limited simulated observations of a typical bright star that shows rotationally broadened spectral lines resolved with a high resolution spectrograph.
We take the parameters for an A6V star of magnitude $V = 4$ observed with an 8\,m telescope using a high resolution spectrograph and assume a signal-to-noise (SNR) of 1200/pixel.
Simulated observations are created by adding white noise at this spectral SNR to the normalized line profiles. 
We assume 21 spectra are taken over a 30 minute period per individual night, and that there are a total of 152 nights of observations.
Lastly, in line with the values for a typical A6V star from \citet{DeMooij17}, we assume an intrinsic line width of $20 \ \rm{km/s} \ \rm{FWHM}$, projected equatorial stellar rotation $v_{\rm{eq}} = 130 \ \rm{km/s}$ (sometimes also referred to as $v \sin i$), V-band limb darkening coefficients for an effective temperature of 8000 K and $\log g = 4.0$ \citep{Claret00} and an intrinsic line depth $A = 0.8$.
We create simulated residual spectra (after median line profile subtraction) at our spectral SNR for the full observation window, and then we apply the following steps to calculate the exoplanet SNR at different stellar positions and exoplanet radii:
\begin{itemize}
    \item An exoplanet with a given radius is injected into one night of spectra with impact parameter $b=0$, at a given radial velocity offset.
    \item The exoplanet signal is calculated by summing up all the flux in $24 \ \rm{km/s}$ (8 pixel wide) bins.
    \item The noise is estimated as the standard deviation of the signals over all the other nights, where no signal was injected. We assume only one planetary transit occurs in the data.
\end{itemize}
This routine is repeated for all nights and with varying the injected planet radius and radial velocity offsets.
A planet is said to be recovered if the SNR $>$ 3.0 for the injected planets location.
The recovery fraction is the number of recovered planets normalised by the total number of nights.
The result is shown in Figure~\ref{fig:wnsen}.
We compare the transiting exoplanet radius $R$ to the radius of Jupiter $R_{\rm{Jup}}$, Saturn $R_{\rm{Sat}}$ and Neptune $R_{\rm{Nep}}$.
Companions with radii $R > R_{\rm{Nep}}$ are fully recovered and radii $R = R_{\rm{Nep}}$ are recovered ${\sim}80\%$ of the time.
For the latter, the recovered fraction is less for larger radial speeds.
This is due to limb darkening, which causes a weaker line profile distortion towards the stellar edges.

\subsection{Period completeness and coverage}
\label{ss:coverage}

\begin{figure}
    \centering
    \includegraphics[width= \columnwidth]{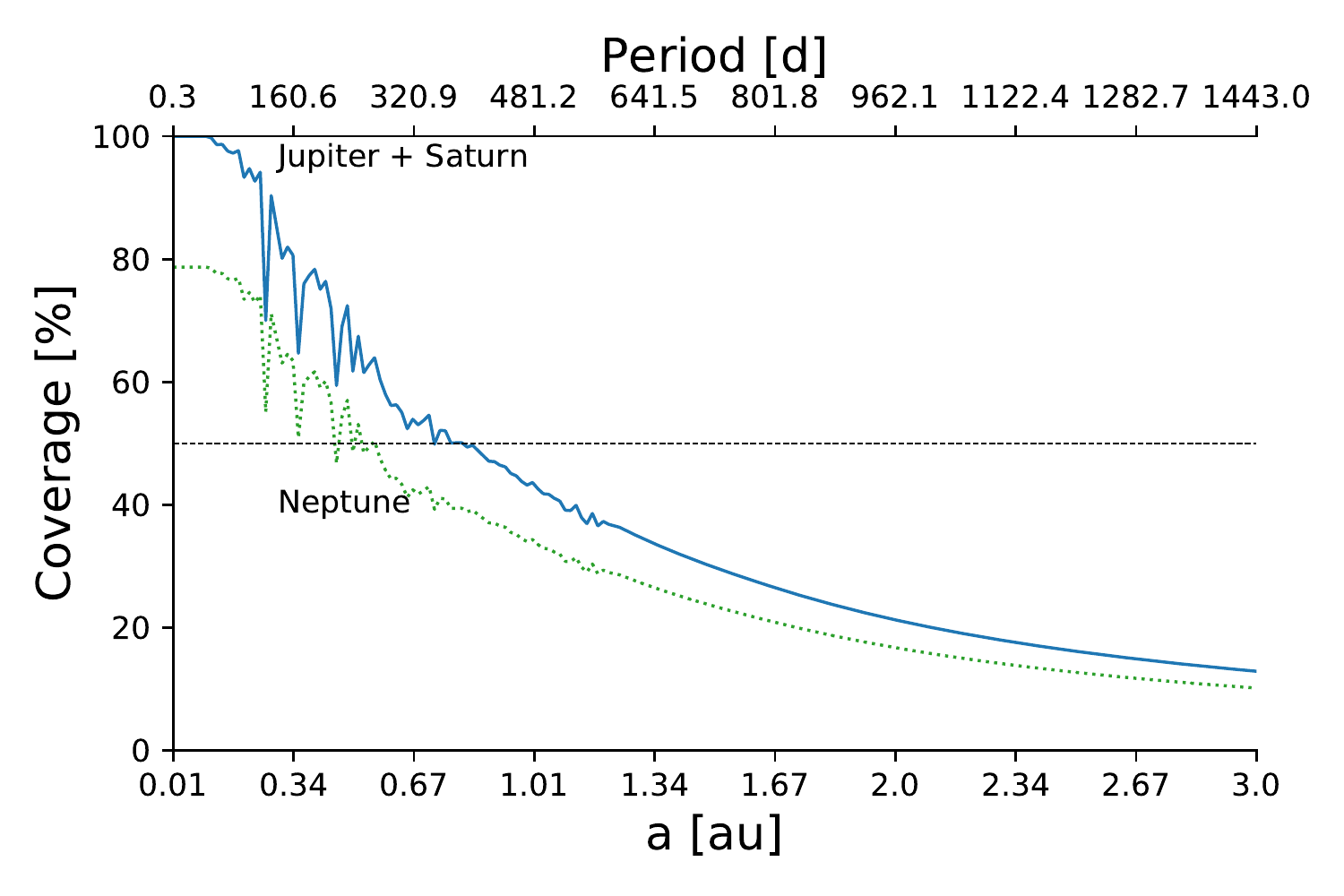}
    \caption{Coverage of our observation window for Jupiter-, Saturn- and Neptune-sized object for our white noise simulation assuming an impact parameter $b = 0$.
    The period completeness equals the Jupiter and Saturn coverage as their sensitivity is 100$\%$.
    For longer periods, the period completeness decreases with decreasing slope. The horizontal dashed line indicates 50$\%$ coverage.}
    \label{fig:coverage_wn}
\end{figure}

Our coverage, Cov($R$, $P$) is the product of the sensitivity, Sen($R$), which is the probability of detecting an exoplanet of size $R$, and the period completeness Com($P$), which is the probability to detect a transit given our observation window.
The sensitivity is averaged over all radial velocity offsets in Figure~\ref{fig:wnsen}.
The period completeness depends only on the observation window and is calculated the following way: 
\begin{itemize}
    \item The transit duration is calculated
    \item The observation window function is convolved with the transit duration.
    \item This convolved window function is folded to the exoplanet period.
    \item The coverage is calculated by taking the ratio of non-zero values over zero values of the period folded convolved window function.
\end{itemize}
The coverage results in this photon shot noise limited case are shown in Figure~\ref{fig:coverage_wn}.
For radii $R \geq R_{\rm{Sat}}$, the sensitivity is 100$\%$, thus the coverage coincides with the period completeness.
Periods $<$13 d ($\sim$0.13 au) are fully complete.
For longer periods the completeness decreases with decreasing slope.
This is due to longer transit durations for longer periods.
Periods ${\sim}1 \ \rm{yr}$ ($\sim$0.8 au) have a 50$\%$ completeness and periods $\sim$1400 d ($\sim$3 au) have a $\sim$10$\%$ completeness.
Small coverage fluctuations are seen for periods $<$1 au.
This is due to the non-uniformity of our window function, which causes overlap of the observation windows when folded for certain periods.
Nonetheless, this effect is negligible, as the resolved gaps have a FWHM of $\sim$0.1 d.
Note that in contrast to photometric light curve transits, we do not need to follow the whole transit from ingress to egress.
Instead, the spectral line profile distortion determines the diameter of the transiting object.

\section{Application to real data: search for transiting planets orbiting \texorpdfstring{$\beta$}{Beta} Pictoris}
\label{s:datasensitivity}

Section~\ref{s:principle} describes an idealized scenario. In this section we apply our method to analyse high-spectral resolution observations of $\beta$ Pictoris obtained in 2017-2018.

\subsection{Observations}

We obtained observations on 160 epochs between April 1, 2017 and April 17, 2018 with the Ultraviolet and Visual Echelle Spectrograph (UVES) at the Very Large Telescope (VLT) in Chile~\citep{Dekker00}.
During each observation the data were obtained simultaneously with the red and the blue arm using the \#2 dichroic and the CD4 and CD2 cross dispersers in the two arms, respectively.
For the wavelength mode the 437+760 mode was selected, resulting in a wavelength coverage of 3760 {\AA} to 4980 {\AA} and 5700 {\AA} to 9450 {\AA} in the blue and red arms, respectively.
Due to the large amount of telluric lines in the red arm, we focus only on data in the blue arm in this paper. We used a 0.3$\arcsec$ slit to obtain the highest possible resolution ($\sim$90,000 before an instrument intervention by the observatory in October 2017, which resulted in an increase in resolving power to 100,000.).
The exposure time for the blue arm was 15 seconds. For the first 10 epochs, we used the 225 kHz readout mode, and obtained 15 exposures per visit; for the remainder we used the 625 kHz mode and obtained 21 exposures per visit.

The data were initially reduced using the UVES pipeline version 5.7.0 via the ESO Reflex \citep{Freudling13}, and we used the spectra before merging of the individual orders for the remainder of our analysis.
After the initial reduction, the spectra were interpolated onto a common wavelength grid and corrected for blaze variations between spectra and nights.
The blaze correction was done on a frame-by-frame and order-by order basis by fitting a low-order polynomial to the ratio of the frame and echelle order being considered and the same echelle order in the reference frame.
For the reference frame, the average spectrum for the sixth visit was used.
After blaze correction, we selected 16 lines across the different orders that visually appeared to be unblended, and cut out a region of $\sim$215 km/s around the apparent line-center.
To improve the SNR ratio and to allow a the individual lines to be combined directly, we binned the data onto a velocity grid with a pixel size of 3 km/s.  
After visual inspection of all nights, we flag 8 nights due to bad data quality, making in total a 152 nights available for further analysis. The flagged nights are (UT): 2017 Apr 8, Sep 11, Oct 5, Oct 8, Oct 9, Oct 10, Oct 13 and 2018 Feb 3. 

\subsection{Stellar pulsations}
\label{ss:sp}
In contrast to our white noise simulations presented in Section~\ref{ss:wnsim}, $\beta$ Pictoris is a $\delta$ Scuti non-radial pulsator \citep{Koen03}, which will have a direct impact on our sensitivity.

\begin{figure}
    \centering
    \begin{subfigure}{ \columnwidth}
        \includegraphics[width= \columnwidth]{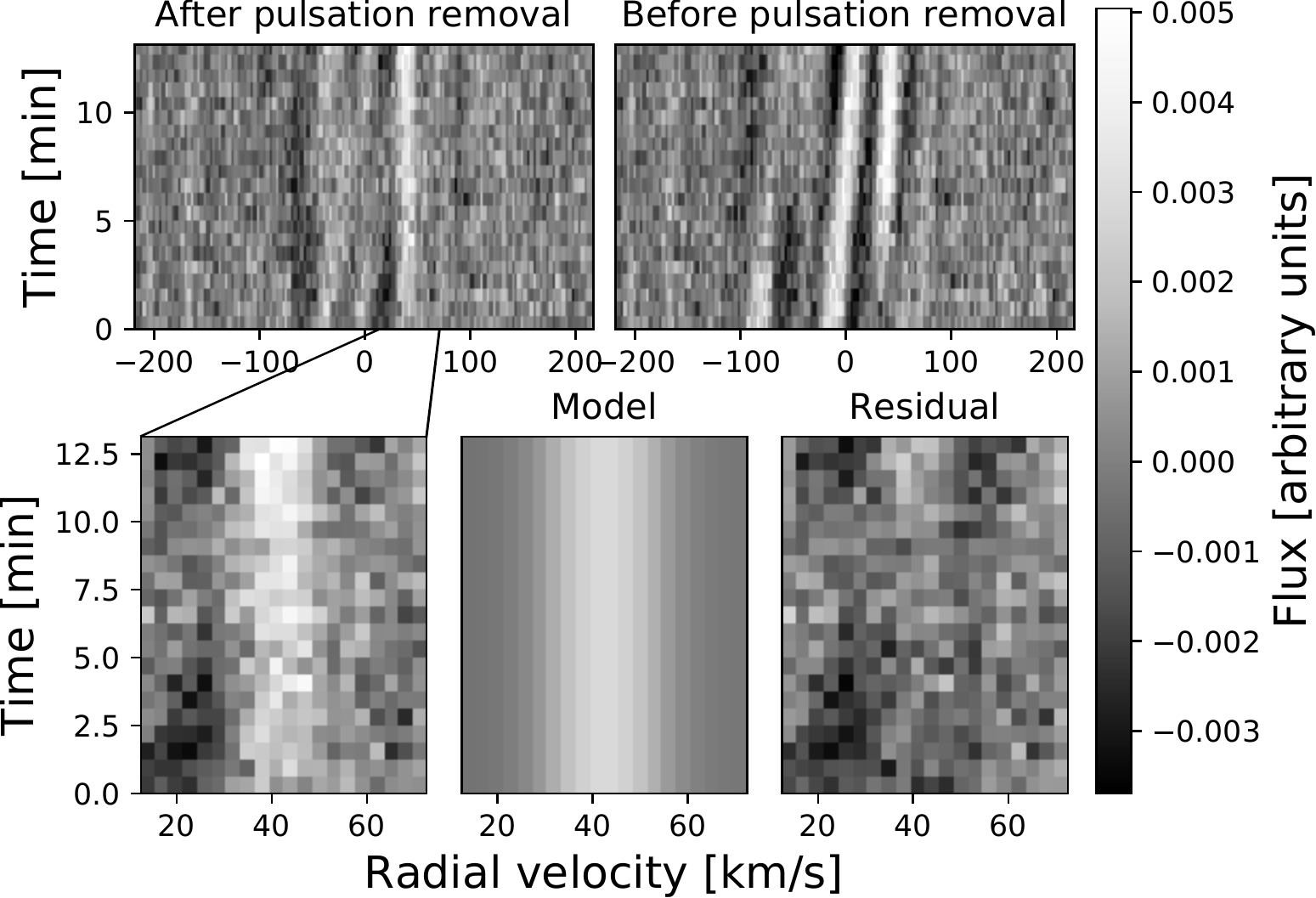}
        \caption{Exoplanet candidate for the night of UT 2017 Sep 8 (JD = 2458004).}
        \label{fig:candidate58004}
    \end{subfigure}
    \begin{subfigure}{ \columnwidth}
        \includegraphics[width= \columnwidth]{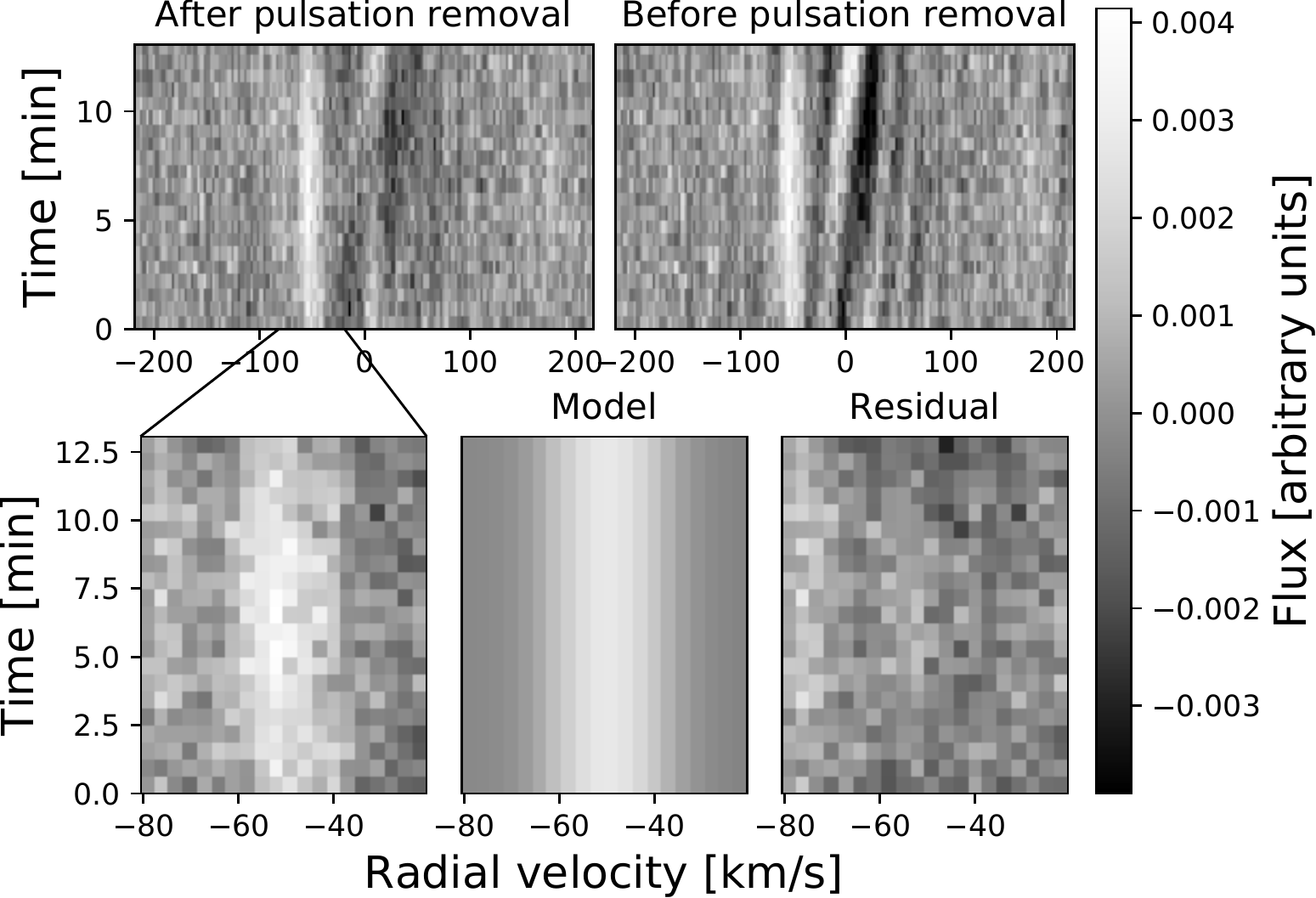}
        \caption{Exoplanet candidate for the night of UT 2017 Dec 11 (JD = 2458098).}
        \label{fig:candidate58098}
    \end{subfigure}
    \caption{Spectral time series of two nights after median line profile subtraction and combining 16 lines.
    The $\delta$ Scuti pulsations can be seen in the top-right boxes and differ in amplitude, shape and slope.
    Two exoplanet candidate signals are boosted after pulsation removal.
    Zooms of the signals, RM models and residuals are shown in the bottom three boxes.}
    \label{fig:candidate}
\end{figure}

For a given set of spectra on a single night, the median stellar profile of all of the other nights are subtracted off of the current night.
The stellar pulsations appear as a quasi-sinusoidal signal in velocity space, which change as a function of time. 
The peak of a given pulsation in velocity space is assumed to vary linearly with observing time.
If the peak amplitudes do not change, they appear as vertical black and white stripes in the residual spectral line time sequence, as seen in the upper right boxes of Figure~\ref{fig:candidate58004} and~\ref{fig:candidate58098}.

The pulsation amplitudes are similar to those expected for $\sim$Jupiter-sized exoplanet signals, thus pulsation removal is required to detect smaller radii companions.
This pulsation removal has been done successfully for broadband time-series photometric observations \citep{Zwintz18, Lous18} and for spectral observations with a companion on a retrograde short-period (1-2 d) orbit \citep{Johnson15, Temple17}.
However, this has not been done before for spectral observations of companions on prograde orbits or on retrograde orbits with $>2$ d periods.
We have developed a method to perform the pulsation removal\footnote{Our pipeline is publicly available at \url{https://github.com/lennartvansluijs/Spectroscopic-Transit-Search}} which we present in this subsection.
We provide the results on the detection limits and coverage, including any residual pulsations, for $\beta$ Pic in Subsection~\ref{ss:senresults}.
The main difficulty is the degeneracy between the stellar pulsations and exoplanet signals in the spectral time series.
A stellar pulsation model using many free parameters fits can overfit an exoplanet signal, whereas using fewer free parameters does not fit the pulsations well when our pulsation model breaks down.
We tried the following approaches: (1) Principle Component Analysis (PCA) on all spectra and removal of the most dominant eigenvectors within the spectra (2) sinusoidal-fitting per night (3) shearing the spectra and applying PCA (4) shearing every night and subtracting off the mean along the time-axis (5) same as (4), but with an extra correction for the contribution of an exoplanet signal.
The last method resulted in the best SNR improvement, and we describe it below.

The radial velocity change during one observation ($\sim$10-20 min observing time) ${\Delta v}_{\rm{obs}}$ due to stellar pulsations is $\sim30-50 \ \mathrm{km/s}$.
The change in radial velocity due to the planet blocking out different velocity strips of the stellar surface $\Delta v_{\rm{obs}}$ of an edge-on and aligned planet for an observation duration $t_{\rm{obs}}$ and transit duration $t_{\rm{transit}}$ is:

\begin{equation}
    {{\Delta v}_{\rm{obs}} = 2v_{\rm{eq}} \ \left ( \frac{t_{\rm{obs}}}{t_{\rm{transit}}} \right ) \qquad (t_{\rm{obs}} \leq t_{\rm{transit}})}.
    \label{eq:deltavrad}
\end{equation}

Therefore a hypothetical planet on a 0.1 au orbit around $\beta$ Pictoris would show a shift of ${\Delta v}_{\rm{obs}} \approx 9 \ \rm{km/s}$ in a typical 30 minute observation.
This number will be even smaller for longer orbital periods, so an exoplanet radial velocity signal is approximated as being constant during the 30 minutes of spectra.
The planet signal appears as a dark vertical stripe in the time series of spectra of 30 minutes.
However, transiting objects on a prograde short period orbits will have slopes aligned with the stellar pulsation signals, making it more difficult to detect them.
On the contrary, such objects on retrograde orbits are easier to detect as they skew from the stellar pulsation signals (as for the Doppler shadow in the HD 15082 (WASP 33) system \citep{Cameron10b}).
The difference in ${\Delta v}_{\rm{obs}}$ between an exoplanet and pulsation signal can be exploited by `shearing' the spectra as shown in Figure~\ref{fig:datareduction}: shifting each spectrum $i$ observed at epoch $t_{i}$ in radial velocity space $\Delta v_i$ proportional to their time difference $\Delta t_i$ with the mid-time epoch of the nightly set of spectra $t_0$.
We define the shearing constant $S$ in $\Delta v_i = S t $ where $t = t_{i} - t_0$.
This aligns most of the stellar pulsations and shears them into vertical lines, aiding their estimation and subsequent modeling and removal.
For the stellar pulsation removal, the following steps are applied as shown in Figure~\ref{fig:datareduction}:
\begin{figure}
    \centering
    \includegraphics[width=  \columnwidth]{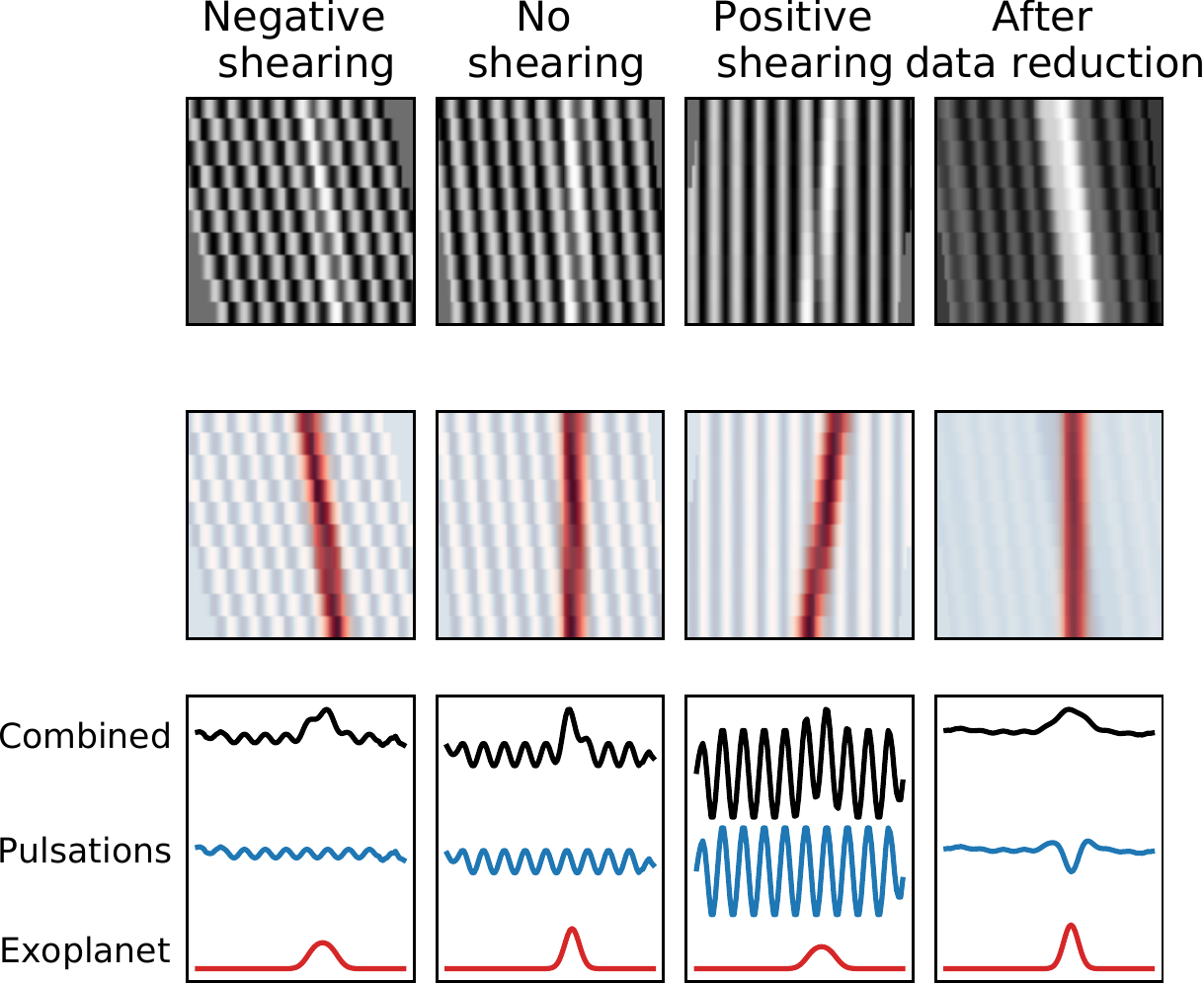}
    \caption{Illustration of our pulsation removal algorithm.
    Black is the observed combined signal, blue the pulsation component (modeled by an inclined sinusoidal signal) and red the exoplanet component (modeled by a Gaussian signal).
    Positive shearing of the spectra aligns the pulsations.
    Due to symmetry, an equal negative shearing estimates the exoplanet signal for the positive shearing.
    Subtraction of the estimated exoplanet and pulsation signal then reverse shearing boosts our exoplanet signal and suppresses pulsations.}
    \label{fig:datareduction}
\end{figure} 

\begin{itemize}
    \item Positive velocity shearing $+S$ is applied for differing values of the shearing constant (3rd column in Figure~\ref{fig:datareduction}).
    \item The exoplanet signal (red) is estimated for all different shears by applying an equal and opposite shearing (1st column).
    \item The exoplanet signal estimates are subtracted off (3rd column minus 1st column).
    \item The pulsations are estimated by the residual of the positive shearing for which the pulsations aligned the best (blue line in 3rd column).
    \item The pulsation estimate is subtracted off the sheared data.
    \item The shearing is reversed ($-S$) and the spectra summed in the time direction.
\end{itemize}
    
The final result for our mock data example is shown in the right-most column of Figure~\ref{fig:datareduction}: a clear SNR improvement with respect to the central column.

\subsection{Sensitivity and coverage}
\label{ss:senresults}
\begin{figure}
    \center
    \includegraphics[width= \columnwidth]{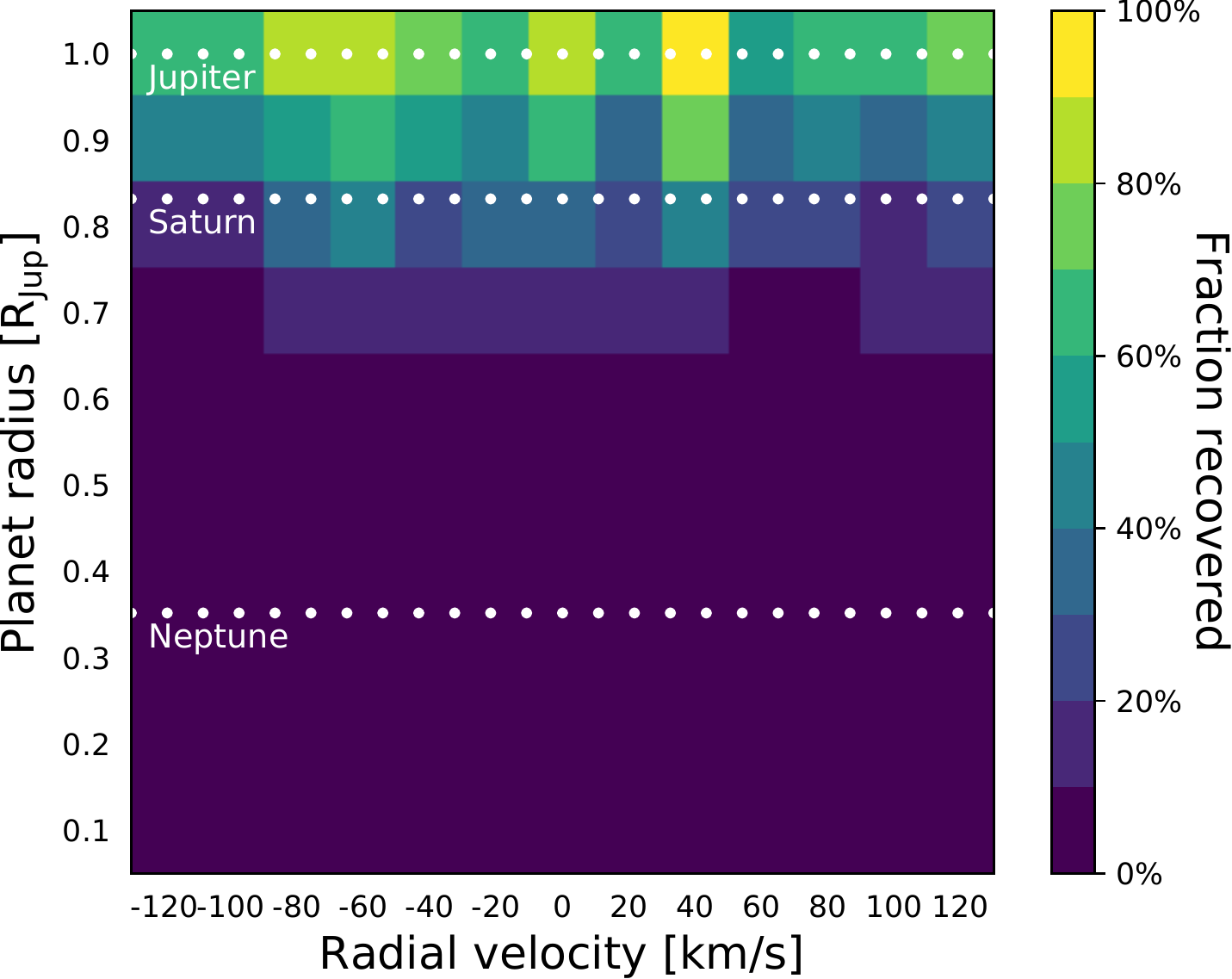}
    \caption{Recovery fraction of 152 transit injections into the $\beta$ Pictoris dataset. %
    Exoplanet signals with a SNR $> 3.0$ above other nights are considered recovered.
    We assume an impact parameter $b = 0$.
    The radii of Jupiter, Saturn and Neptune are indicated by the white dashed lines.
    As $v_{\rm{eq}} = 130 \ \rm{km/s}$, radial velocities outside of our range are (close-to) zero.
    Companions with radii $R = R_{\rm{Jup}}$ are almost fully recovered and a large fraction with radii $R = R_{\rm{Sat}}$ are recovered.
    The effects of stellar pulsation are seen by the fluctuations of the recovery fraction as a function of radial velocity.
    }
    \label{fig:betapicsen}
\end{figure}

\begin{figure}
    \centering
    \includegraphics[width= \columnwidth]{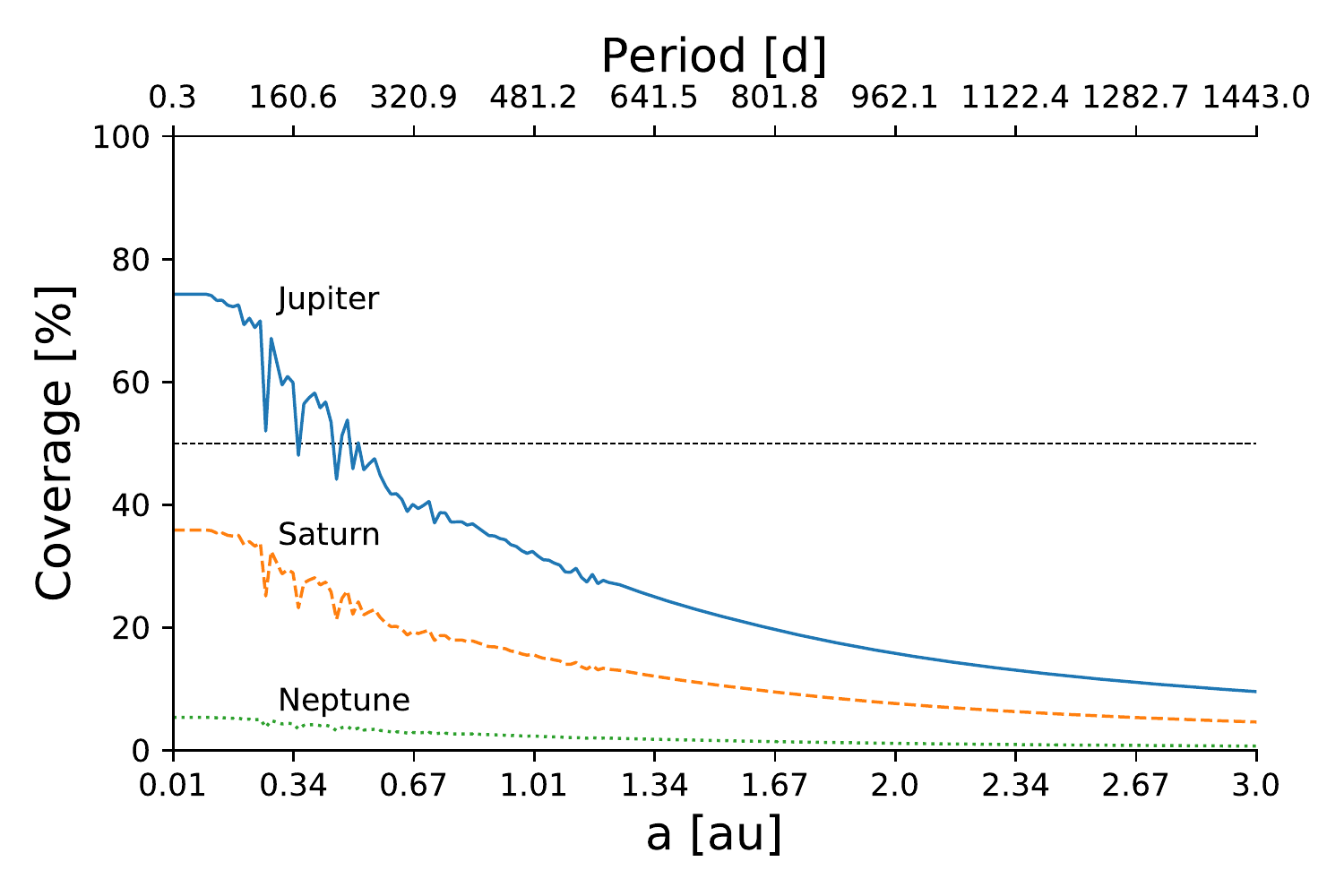}
    \caption{Coverage for our observation window for Jupiter-, Saturn- and Neptune-sized objects in the stellar pulsation-limited case for an impact parameter $b = 0$.
    For longer periods, the coverage drops with decreasing slope.
    The horizontal dashed line indicates 50$\%$ coverage.}
    \label{fig:coverage_betapic}
\end{figure}

Residual spectra for all observations of $\beta$ Pictoris are created the following way:
\begin{itemize}
    \item The median line profile of all nights for all lines is calculated and used as a reference line profile.
    \item All line profiles are normalized with respect to the reference line profile.
    \item The reference line profiles are subtracted off.
\end{itemize}
The same planet injection routine is applied, combined with our pulsation removal routine.
The results are shown in Figure~\ref{fig:betapicsen}.
Companions with $R = R_{\rm{Jup}}$ are recovered for ${\sim}74 \%$ of all nights.
The recovered fraction of $R = R_{\rm{Sat}}$ companions is ${\sim}36 \%$.
As the stellar and observational parameters described in Section~\ref{ss:wnsim} match with $\beta$ Pictoris, we can directly compare Figures~\ref{fig:wnsen} and ~\ref{fig:betapicsen}.
This shows pulsations limit our sensitivity to Saturn- to Neptune-sized objects.
Using these sensitivity limits we are able to calculate coverage limits as described in Subsection~\ref{ss:coverage}, with the results shown in Figure~\ref{fig:coverage_betapic}.
Neptune-sized objects in the photon noise limit and Jupiter-sized objects in the pulsation limit have almost equal coverage, which shows the direct impact of the stellar pulsations on our sensitivity.

\subsection{Comparison with previous surveys}

One previous exoplanet transit search around $\beta$ Pictoris has been conducted using data from the BRITE-Constellation nanosatellite BRITE-Heweliusz \citep{Lous18}.
Their coverage for periods ${\leq}15 \ \mathrm{d}$ is similar to ours for $R = R_{\rm{Jup}}$ and slightly better for $R = R_{\rm{Sat}}$.
For periods ${\geq}15 \ \mathrm{d}$ our coverage is better.
This is partially due to the difference in completeness: 78 nights for \citet{Lous18} and 152 nights for this work.
Nonetheless, for a conventional transit survey detection, coverage of the full transit is required.
Our proposed method does not require full transit coverage, as coverage of only part of the transit will already reveal a spectral line profile distortion.
This is an advantage of our method over conventional transit surveys.
Additionally, exoplanet radial velocity studies have been done by \citet{Lagrange13} and \citet{Lagrange18b}.
Following \citet{Lous18}, the mass upper limits at different orbital periods can be converted into radii using {\sc Forecaster}\footnote{Available at \url{https://github.com/chenjj2/forecaster}} \citep{Chen16}.
Compared to \citet{Lagrange13}, the sensitivity increases significantly for $R < R_{\rm{Jup}}$ objects at smaller orbital periods.
One major result of \citet{Lagrange18b} is the exclusion of companions more massive than 3 $M_{\rm{Jup}}$ closer than 1 au and further than 10 au, with a 90$\%$ probability.
Even in the fully recovered case, we have a coverage of $>45\%$ at 1 au (see Figure~\ref{fig:coverage_wn}), less than 90$\%$.
For a very massive transiting companion $< 1 \ \rm{au}$ and at very long orbital periods ($\ga 200 \ \rm{d}$) the radial velocity search has greater sensitivity.
For the latter, as geometrical transit probabilities are very small in this regime, it is unlikely any future transit survey will be competitive in this regime.

\subsection{Candidates}
\begin{figure}
    \centering
    \includegraphics[width= \columnwidth]{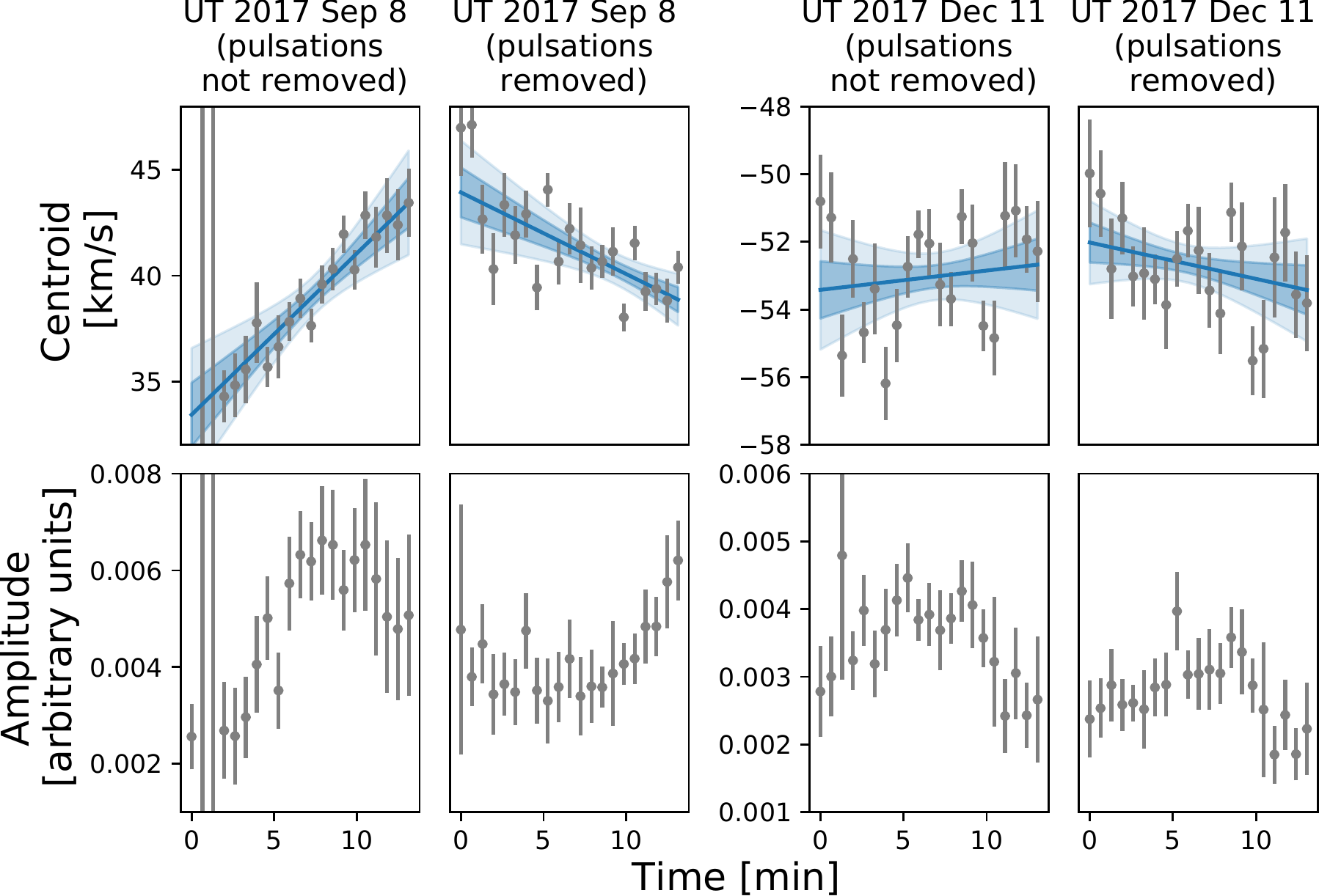}
    \caption{Gaussian with variable background best-fits to the spectral time series of the two candidates in Figure~\ref{fig:candidate}. This has been done both before and after stellar pulsation removal.
    The best linear fit of the centroid time series and corresponding 1$\sigma$- and 2$\sigma$-confidence intervals are shown in blue and light blue. 
    The bottom row shows the best-fit amplitude time series.}
    \label{fig:gaussianfit}
\end{figure}

A similar analysis search in our data results in 12 nights containing a signal with a SNR > 3.0.
Visual inspection shows most of these are due to strong outlier pixels, strong stellar pulsations or bad data quality.
Two candidates remain (see Figure~\ref{fig:candidate}), for both, a $R = R_{\rm{Jup}}$ at an edge-on orbit object fits the signal well.
Gaussian profiles with variable background are fitted using Levenberg-Marquardt minimization to each spectral time series, both before and after stellar pulsation removal, to obtain an amplitude- and centroid time series.
A linear fit to the latter constraints the Doppler shadow's slope.
The results are shown in Figure~\ref{fig:gaussianfit}.
After pulsation removal, we find $\Delta v_{\rm{rad}} = -5.1 \pm 1.6 \ \rm{km/s}$ (2017 Sep 8) and $\Delta v_{\rm{rad}} = -1.4 \pm 1.2 \ \rm{km/s}$ (2017 Dec 11), both suggesting retrograde orbits.
Orbital period limits are calculated using Equation~\ref{eq:deltavrad}.
We find $P = 0.2_{-0.1}^{+0.3} \ \mathrm{yr}$ (2017 Sep 8) and $P = 6.0_{-4.8}^{+\inf} \ \mathrm{yr}$ (2017 Dec 11).
For the candidate of 2017 Sep 8 a detection is plausible based on the period lower limit, for which we have ${\ga}75 \%$ coverage.
However, the amplitude time series shows a strong increase in amplitude over the observational duration, which is expected for the stellar pulsations with short periods, but not for a transiting companion.
This is supported by the radial velocity change before pulsation removal of $\Delta v_{\rm{rad}} = 10 \pm 2.5 \ \rm{km/s}$, much closer to typical stellar pulsation values.
For the candidate of 2017 Dec 11 it is plausible with the period lower limit, for which we have ${\ga}35 \%$ coverage.
The slope in the velocity during the observations is consistent both before and after pulsation removal with an exoplanet signal, however the amplitude time series shows a significant variation, suggesting that the signal is not planetary in nature.
Consequently, we conclude both signals are unlikely due to a transiting exoplanet.
The retrieval of these two signals demonstrate the capability of our algorithm to retrieve exoplanet-like signals and the subsequent analysis shows it is possible to identify false-positives from their slope and time variation.
We find these results very encouraging, but this also demonstrates the detection of at least three transits will be required to confirm the planetary nature of the signals in the future.

\section{Discussion}
\label{s:discussion}
\begin{figure}
    \centering
    \includegraphics[width= \columnwidth]{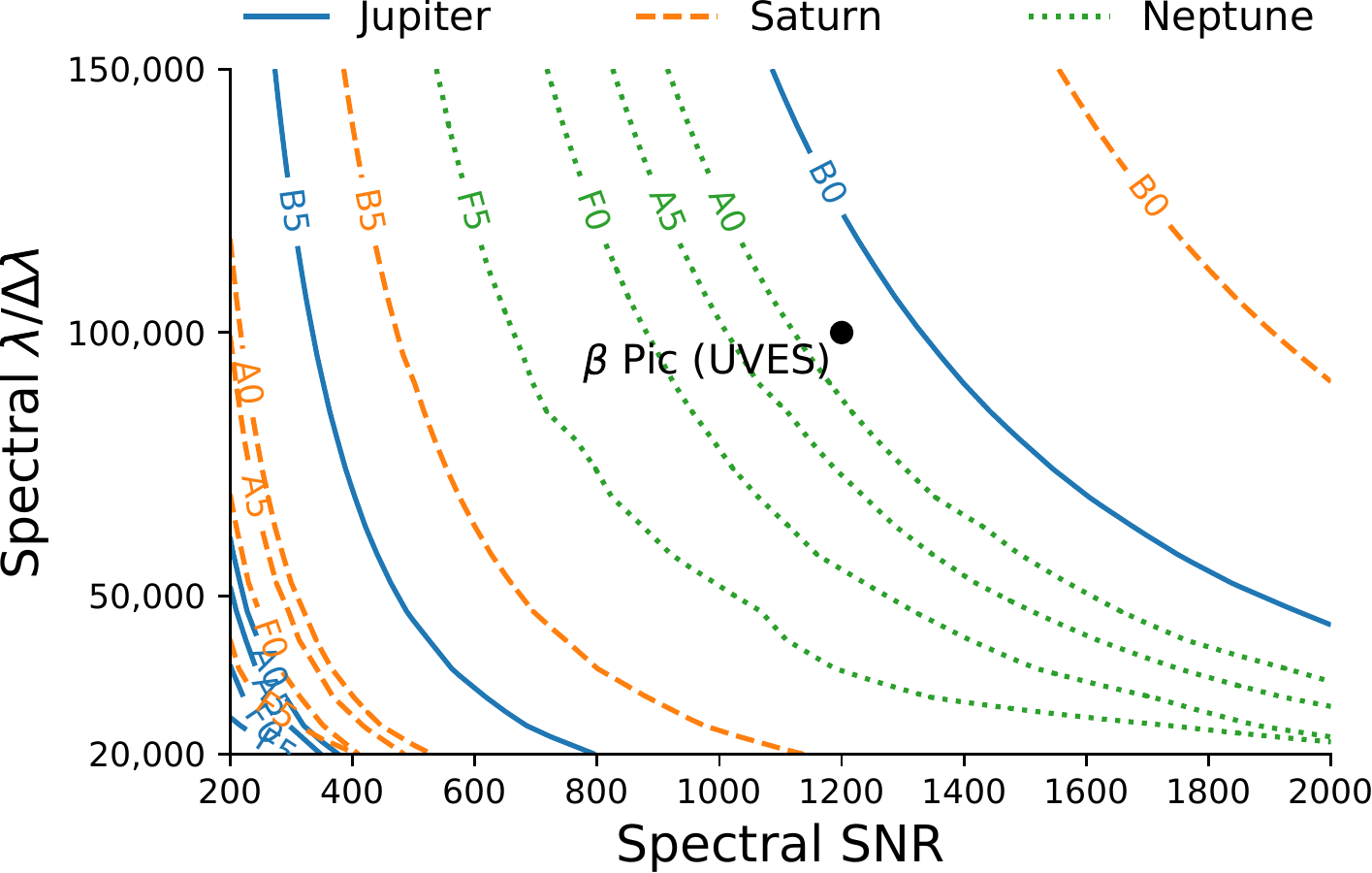}
    \caption{Contours for different BAF-spectral type stars for which a Jupiter- Saturn- and Neptune-sized object can be detected at a SNR > 3.0 following a survey of 20 minutes per night for 152 nights.
    For a higher spectral SNR, smaller objects can be detected.
    If the spectral resolution is too low, the spectral line cannot be resolved anymore.
    For an instrumental setup to the right of a contour, an exoplanet detection is feasible for photon-shot-noise limited observations.
    The instrumental setup used for our observations of $\beta$ Pictoris with UVES on the VLT is indicated by the scatter point.}
    \label{fig:spectralsurvey}
\end{figure}

In the previous sections we have shown the principle of the method for a typical bright star, $\beta$ Pictoris.
In this section we discuss the possibility of applying our method to other bright stars in the sky.
Therefore, we extend our analysis to a broader range of spectral types and instrumental parameters.
According to the SIMBAD database \citep{Wenger00}, there are 512 stars with  $V$ < 4 observed by Hipparcos \citep{Perryman97}.
Among these, a majority of 272 stars are BAF-spectral types.
These are expected to rotate fast enough to resolve the planet Doppler shadow, but have stellar radii small enough to detect exoplanets.
For these spectral types we simulate line profiles of planets transiting in front of the stellar center.
These profiles are calculated on an over-sampled velocity grid of 1\,km/s steps.
We adopt the stellar radii and temperatures by \cite{Mamajek13}\footnote{\url{http://www.pas.rochester.edu/~emamajek/EEM_dwarf_UBVIJHK_colors_Teff.txt}} as shown in Table~\ref{tab:stellarparams}.
\begin{table}
	\centering
	\begin{tabular}{l c c r}
{Spectral type} & {Radius [$R_{\sun}$]} & {$v_{\rm{eq}}$ [km/s]} & {T{\rm{eff}} [K]} \\ \hline
B0 & 7.53 & 350 & 31500 \\
B5 & 3.40 & 330 & 15700 \\
A0 & 2.09 & 310 & 9700 \\
A5 & 1.94 & 290 & 8080 \\
F0 & 1.79 & 170 & 7220 \\
F5 & 1.46 & 40 & 6510 \\
    \end{tabular}
  	\caption{Stellar parameters used for the white noise simulations of different stellar spectral types.}
  	\label{tab:stellarparams}
\end{table}
A future survey would likely aim to search for companions around the fastest rotating stars first, as (1) they are more likely to have an edge-on inclination thus a transiting companion and (2) it is easier to resolve the spectral line profiles.
Therefore, we follow the upper bounds on the $v_{\rm{eq}}$ values for each spectral type estimated from Figure~18.21 from \cite{Gray05} (see Table~\ref{tab:stellarparams}).
The intrinsic line width $v_{\rm{int}}$ due to thermal broadening is described by the Maxwell-Boltzmann distribution $v_{\rm{int}} \propto \sqrt{T}$.
Using the temperatures in Table~\ref{tab:stellarparams} a simple scaling relation is adopted to estimate $v_{\rm{int}}$ for all spectral types where we benchmark at $v_{\rm{int}} = 20 \ \rm{km/s}$ and $T = 8000 \ \rm{K}$.
These over-sampled line profiles are convolved and binned to an instrumental spectral resolution $\lambda / \Delta \lambda$.
The planet signal is measured as the sum of all points where the signal is above zero in the out-of-transit subtracted line-profile.
Assuming an instrumental spectral SNR per resolution element, we calculate the SNR of the exoplanet signal.
The results are shown in Figure~\ref{fig:spectralsurvey}.
The contours show the SNR > 3.0 limits for the different spectral types.
For larger stellar radii, the amplitude of the exoplanet signal is smaller.
For larger rotational velocities, a lower resolution is required to resolve the exoplanet signal.
This effect is most prominent for the F5-spectral type, which has the lowest $v_{\rm{eq}}$-value.
For stars with spectral resolutions and SNRs to the right of the line it is feasible to detect a companion of the specified size.
Consistent with our previous result, the $\beta$ Pictoris (an A6 star) observations are on the right-side of the A5 star $R = R_{\rm{Nep}}$ contour, as we are sensitive to $R = R_{\rm{Nep}}$ companions in the photon-shot-noise limited case.

As already seen for $\beta$ Pictoris, stellar activity has a direct impact on our sensitivity.
Most late A- and F-spectral types will suffer the same limitations as they are within the HR-diagram's instability strip \citep{Gautschy96}.
For these objects the stellar pulsation removal procedure described in this work could be applied. For late A and B-stars hotter than $\beta$ Pictoris, the prospects are better, as these are outside of the instability strip and also will not have starspots.
For stars cooler than $\beta$ Pictoris, starspots could be a problem, as they would show up as similar distortions of the line profile \citep[e.g. for $\alpha$ Cen B][]{Thompson17}.
However, for these stars (a) we will know whether they are active (b) we will see the modulation with the rotation period of the star and (c) view differing impact of the stellar activity between lines, while the planet's signal will be the same for all lines \citep{Dumusque18}.
Currently, utilizing UVES on the VLT is the only way to get a sufficiently large sample of high SNR spectra relatively easy.
An advantage of observing bright stars such as $\beta$ Pictoris is that it can be done even during twilight and thus makes optimal use of the telescope. 
Nonetheless, UVES has not been designed with the aim to survey the brightest $V < 4$ stars in the sky.
Firstly, due to the narrow slit width required to get the high spectral resolution (0.3$\arcsec$), in median seeing the slitlosses can be a factor of $\sim$4.
Secondly, there are large overhead losses as we integrate for $\sim$15 s, but have to wait $\sim$45 s before taking the next science image.
The relative overheads will increase even further for brighter ($V < 4$)  stars, resulting in a further reduced efficiency. Lastly, in this work we combined 16 stellar lines, however, in most cases we can use other techniques, including Least-Squares Deconvolution \citep[LSD; e.g.][]{DonatiEtAl97}, to combine a large number of lines and further improve the SNR.
Therefore, we expect a $\sim$1-2 m telescope optimized to observe the brightest ($V < 4$) stars in the sky, could achieve a similar spectral SNR to our VLT observations of $\beta$~Pic in the same amount of on-sky time.
One relatively affordable option would be to refurbish an underutilised 1-2m telescope at an observatory where the seeing can be on the order of one or two arcseconds, although an array of newly constructed telescopes is also an option. Each telescope could observe several dozen stars per night and continue the all sky survey throughout the year.
An optimistic back-of-the-envelope calculation of the expected number of detectable transiting companions around a $V < 4$ star for such a survey find it is $\sim$0.6: the occurrence of transiting $R = R_{\rm{Jup}}$ objects is about one-in-a-thousand \citep{Fressin13} and we estimate a detection feasible around the 272 $V < 4$ B, A \& F stars, respectively for transiting $R = R_{\rm{Nep}}$ companions it is about two-in-a-thousand and we estimate a detection feasible around the 139 $V < 4$ A \& F stars. However, we emphasize this estimate does require improvements on the reduction of stellar phenomena such as stellar pulsations and starspots in the future, especially for the A and F stars.

\section{Conclusions}
\label{s:conclusions}
In this work we demonstrate the RM effect can be used not only to characterize exoplanetary systems, but to also be used for blind spectroscopic transit searches around the brightest rapidly rotating stars in the sky that are challenging to calibrate with reference star observations.
This method is:
\begin{itemize}
    \item independent of reference stars used for conventional broadband transit surveys and works especially well for strong rotationally broadened stars for which radial velocity measurements are difficult.
    \item simulated for observations of a typical bright $V \sim 4$ star at $R \sim 100,000$, for which we show we are sensitive to Neptune-sized objects if the data is photon-shot-noise limited.
    \item applied to a case study, $\beta$ Pictoris, where the ambiguity between stellar pulsations and exoplanet Doppler shadows constrain our sensitivity to Jupiter-sized objects. However, after our pulsation removal procedure we are sensitive to Saturn-sized objects. 
\end{itemize}
These results are currently the strongest constraints on Jupiter-sized transiting companions around $\beta$ Pictoris for periods of 15-200 $\mathrm{d}$ with $>$50$\%$ coverage.
We have considered it feasible to setup a campaign deploying a set of 1-2 m telescopes equipped with a high resolution spectrograph to monitor the brightest stars ($V<4$) aiming to find the brightest star in the sky with a transiting exoplanet.
 This is a high-risk endeavour, but with the potential of a tremendous scientific reward: the discovery of the brightest star in the sky for detailed exoplanet atmospheric characterization and modelling through transmission spectroscopy.

\begin{acknowledgements}
{We are thankful to the Leiden exoplanet group members for the fruitful discussions and their supportive criticism that improved the quality of this work. We also like to thank Remko Stuik for sharing his thoughts on the interpretation of the candidate signals.
This research made use of the Python packages {\tt Astropy} \citep{Astropy18}, {\tt SciPy} \citep{Jones01}, {\tt NumPy} \citep{Colbert11, Oliphant15}, {\tt Matplotlib} \citep{Hunter07}, {\tt LMFIT} \citep{Newville15} and we thank all of their contributors for making their software publicly available.
Part of this research was carried out at the Jet Propulsion Laboratory, California Institute of Technology, under a contract with the National Aeronautics and Space Administration.
}
\end{acknowledgements}

\bibliographystyle{aa} 
\bibliography{bibliography} 

\end{document}